\documentclass[aps,pra,twocolumn,superscriptaddress]{revtex4}
\usepackage{amsmath,epsfig,mathrsfs,graphicx,amssymb}
\def\tr{\mathop{\rm tr}}

\def\be#1\ee{\begin{equation}#1\end{equation}}

\def\nA{\displaystyle{\not}A}
\def\hnA{\displaystyle{\not}\hat{A}}
\def\hnAB{\hat{\displaystyle{\not}A\displaystyle{\not}B}}
\def\nB{\displaystyle{\not}B}
\def\naa{\displaystyle{\not}1}
\def\nbb{\displaystyle{\not}2}
\def\nal{\displaystyle{\not}\alpha}

\newcommand{\ba}{\begin{eqnarray} }
\newcommand{\ea}{\end{eqnarray} }

\begin{document}

\title{Testing locality and noncontextuality with the lowest moments}

\author{Adam Bednorz}
\email{Adam.Bednorz@fuw.edu.pl}
\affiliation{Faculty of Physics, University of Warsaw, Ho\.za 69, PL-00681 Warsaw, Poland}
\author{Witold Bednorz}
\email{wbednorz@mimuw.edu.pl}
\affiliation{Faculty of Mathematics, Informatics, and Mechanics,
University of Warsaw, Banacha 2,
02-097 Warsaw, Poland}
\author{Wolfgang Belzig}
\affiliation{Fachbereich Physik, Universit{\"a}t Konstanz, D-78457 Konstanz, Germany}
\date{\today}

\begin{abstract}

The quest for fundamental test of quantum mechanics is an ongoing effort. We here address the question of what are the lowest possible moments needed to prove quantum nonlocality and noncontextuality without any further assumption -- in particular without the often assumed dichotomy.
We first show that second order correlations can always be explained by a classical noncontextual local-hidden-variable theory. Similar third-order correlations also cannot violate classical inequalities in general, except for a special state-dependent noncontextuality.
However, we show that fourth-order correlations can violate locality and state-independent noncontextuality. Finally we obtain a fourth-order continuous-variable Bell inequality for position and momentum, which can be violated and might be useful in Bell tests closing all loopholes simultaneously.
\end{abstract}

\maketitle
\section{Introduction}
 Certain quantum correlations cannot be reproduced by any classical local-hidden-variable (LHV) theory, as they violate e.g. the Bell inequalities 
for correlations of results of measurements by separate observers\cite{bell}.
The Bell test must be performed
under the following conditions: (i) dichotomy
of the measurement outcomes or at least some restricted set of outcomes in some generalizations
\cite{cglmp}, (ii)
freedom of choice of the measured observables \cite{will}, and (iii) a shorter time of
the choice and measurement of the observable than the
communication time between the observers.
Despite considerable experimental effort \cite{bellex},
the violation has not yet been confirmed conclusively, due to
several loopholes \cite{loop}. The loopholes reflect the fact that the experiments have not 
fully satisfied all the conditions (i-iii) simultaneously.
In fact, the Bell test is stronger than the entanglement criterion, viz. the nonseparability of states \cite{ent}. The latter 
assumes already a quantum mechanical framework (e.g. an appropriate Hilbert space), while the former is formulated classically.
The loophole-free violation of a Bell inequality -- not just the existence of entanglement -- is also necessary to prove the absolute security of quantum cryptography \cite{gis06}.

Nonclassical behavior of quantum correlations can appear also as a violation of noncontextuality. Noncontextuality means that
the outcomes of experiments do not depend on the detectors' settings so that there is a common underlying probability for the results of all possible settings while the accessible correlations correspond to commuting sets of observables. The Kochen-Specker theorem ingeniously shows that noncontextuality contradicts quantum mechanics \cite{kst},
Noncontextuality is testable in realistic setups \cite{kst2}.
In contrast to noncontextuality,  Bell-type tests of nonlocality without further assumptions must exclude also \emph{contextual} LHV models
as correlations of outcomes for different settings are not simultaneously experimentally accessible for a single observer,
even if they accidentally commute.
Moreover, noncontextuality may be violated for an arbitrary localized state (state-independent noncontextuality \cite{stin}) while Bell-type tests make 
sense only for nonlocally entangled states. If a Bell-type inequality is violated then
state-dependent noncontextuality is violated, too, but not vice-versa.

As the Bell and noncontextual inequalities are often restricted to dichotomic outcomes, e.g. $A=\pm 1$,
generalizations have been investigated, including the many-outcome case \cite{cglmp}.
Recently, Cavalcanti, Foster, Reid and Drummond (CFRD) \cite{cfrd} proposed a way to relax the constraint of 
dichotomy, allowing any unconstrained real value. 
CFRD constructed a particularly simple class of inequalities holding classically, while seemingly
vulnerable by quantum mechanics. The inequalities involve $n$th moments $\langle A^{n-l-m}B^lC^m\rangle$ of observables $A$, $B$, $C$, and nonnegative integers $l,m$ and $n-l-m$, where in general the higher $n$ is, the greater the chances to violate the corresponding CFRD inequality.
On a practical level, measuring higher moments or making binning is not a problem if the statistics consists of isolated peaks. However, in many experiments, especially in
condensed matter \cite{three}, the interesting information is masked by 
large classical noise.  This noise then dominates the signal and makes the binning unable to retrieve the underlying quantum statistics, which is accessible only by measuring moments and subsequent deconvolution.

In this paper we ask which are the lowest possible moments to show nonclassicality and systematically investigate whether second-, third- or fourth-order correlations are sufficient to exclude LHV theories.  We first show that second-order inequalities cannot be violated  at all
because of the so-called weak positivity \cite{bb11} -- a simple classical construction of a probability reproducing all
second-order correlations. Note that the standard Bell inequalities \cite{bell} require  \textit{experimental} verification of the dichotomy $A^2=1$, which means e.~g.
showing that $\langle(A^2-1)^2\rangle=0$ by measuring the corresponding fourth-order correlator or applying binning (in which case the correlator is obviously zero). Hence, operationally a standard Bell test is of at least fourth order -- not second, as it may appear from the Bell inequalities \cite{bell} alone. We emphasize that binning is useless, if the signal is masked by classical noise. 
The proposed Bell-type tests in condensed matter based on second order correlations \cite{entsol,entrev,heiblum} require an additional assumption of a dichotomous interpretation of the measurement results, which is in general experimentally unverified and does not allow entanglement to be identified unambiguously. Next we will show, that Bell-type tests for third moments with standard, projective measurements are not possible. 
Nevertheless, third moments can violate noncontextuality but only for a positive semidefinite correlation matrix and special states. 
Our main result is to show that generally fourth-order correlators are sufficient to violate state-independent noncontextuality and a  Bell-type inequality which can be violated by correlation of position and momentum in a special entangled state. State-independent noncontextuality can be violated by a fourth-moment generalization of the Mermin-Peres square \cite{merper}. Our results for the gradual possibilities of excluding LHV models under different conditions are summarized in Table \ref{tab1}. 


Comparing to the previous research, note that
the CFRD inequalities are the only known Bell-type inequalities scalable with $A\to\lambda A$, $B\to\mu B$ and so on for more observers.
Unfortunately, the original example for a violation involved 20th-order correlators and 10 observers \cite{cfrd}, but 
was later reduced to  6th order and 3 observers \cite{salles,qhe} for Greenberger-Horne-Zeilinger states
\cite{ghz}. On the other hand, 
the CFRD inequality with 4th moments cannot be violated at all, which has been shown
for spins \cite{pro1}, quadratures \cite{pro2}, generalized to 8 settings and proved for separable states \cite{schvo},
and finally proved for all states \cite{salles} (we show an alternative proof in Appendix E). 

The paper is organized as follows. We start with a general description of tests of contextuality and locality.
Then we show that second moments are insufficient to violate locality and noncontextuality. Next, we show that third moments are enough only to show state-dependent contextuality. In the last part we discuss fourth moments, which allow violation of state-independent noncontextuality and locality. The violation of locality is possible with moments of positions/momenta (quadratures).

\begin{table}
\begin{tabular}{l|llll}
\hline
Noncontextuality &Yes &Yes& No\\
\hline
State independent&No&Yes&No\\
\hline
Maximal moments& \multicolumn{3}{c}{LHV excluded?}\\
\hline\hline
2nd &No&No&No\\
\hline
3rd &Yes&No&No\\
\hline
4th &Yes&Yes&Yes\\
\hline
\end{tabular}
\caption{Summary of the feasibility of moment-based tests of LHV theories depending on the conditions: a) contextuality or noncontextuality and b) special or arbitrary input state. The entries answer the questions: Are correlations with moments up to the given order not explicable by a joint positive probability?}
\label{tab1}
\end{table}



\section{Test of local-hidden-variable models} 
\begin{figure}
\includegraphics[scale=.5]{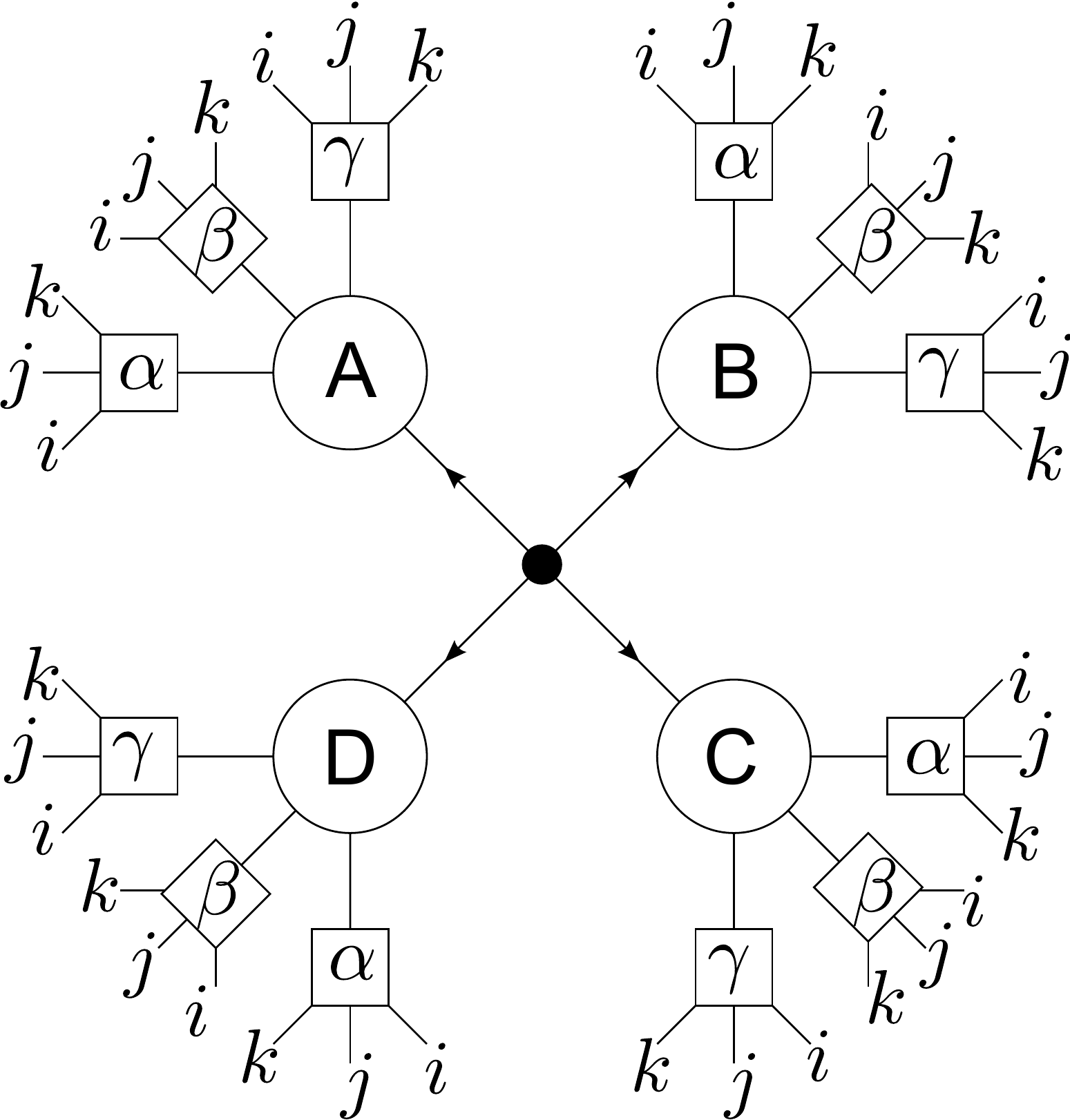}
\caption{The general test of local realism. Here we have four observers, Alice, Bob, Charlie and David. Everybody is free to choose between three different
settings, $\alpha$, $\beta$ and $\gamma$ and finally they can measure three real, continuous outcomes, 
e.g. $A_{\alpha i}$. The picture can be generalized to arbitrary numbers of observers, settings and outcomes.}\label{bellset}
\end{figure}

Let us adopt the Bell framework, depicted in Fig.~\ref{bellset}. Suppose Alice, Bob, Charlie, etc. are separate observers 
that can perform measurements  on a possibly entangled state, which is described by an initial density matrix $\hat{\rho}$.  
Every observer $X=A,B,C,\dots$ is free to prepare one of several settings of their own detector ($\alpha=1,2,\dots$). 
For each setting, one can measure multiple real-valued observables
(numbered $i=1,2,3,\dots$) so that the measurement of $\hat{X}_{\alpha i}$ gives a real number $X_{\alpha i}$ 
The projection postulate gives the quantum prediction for correlations,
$\langle O_1\cdots O_n\rangle=\mathrm{Tr}\hat{\rho}\hat{O}_1\cdots \hat{O}_n$
for commuting observables $\hat{O}_k$. 
The observables measured by different observers and by one observer $\hat{X}_{\alpha i}$ for a given setting have to commute, viz. 
$[\hat{X}_{\alpha i},\hat{Y}_{\beta j}]=[\hat{X}_{\alpha i},\hat{X}_{\alpha j}]=0$. 
The observables for one observer but different settings, $\hat{X}_{\alpha i}$ and $\hat{X}_{\beta j}$ for $\alpha\neq\beta$, may be noncommuting but may also accidentally commute
or even be equal.  A LHV model assumes the existence of a joint positive-definite probability distribution of all possible outcomes $\rho(\{X_{\alpha i}\})$ that reproduces quantum correlations for a given setting.
If the accidental equality between observables for different settings, $\hat{X}_{\alpha i}=\hat{X}_{\beta j}$, imposes the constraint 
${X}_{\alpha i}\equiv{X}_{\beta j}$ in $\rho$,  the LHV model is called \emph{noncontextual}. 
A single observer suffices to test such LHV as noncontextuality  is anyway an experimentally unverifiable assumption --
the observer  cannot measure simultaneously at two different settings. 
In contrast to noncontextuality, the \emph{locality} test must allow \emph{contextuality}: that even if $\hat{X}_{\alpha i}=\hat{X}_{\beta j}$ 
($\alpha\neq \beta$)
then ${X}_{\alpha i}\neq{X}_{\beta j}$ is still possible.
The choices of the settings and measurements are required to be fast enough to
prevent any communication between observers. Then $\rho$ cannot be altered by the choice of the observable. 
Noncontextual and local LHVs can be ruled out by tests with discrete outcomes \cite{bell,kst}. In moment-based tests only a finite number of cross correlations are compared with LHV. Our aim is to find the lowest moments showing nonclassical
behavior of quantum correlations.

\section{Weak positivity} 
For a moment all observables, commuting or not, will be denoted by $\hat{X}_i$.
Let us recall the simple proof that first- and second-order
correlations functions can be always reproduced classically \cite{bb11}.
To see this, consider a real symmetric correlation matrix
\begin{equation}
\mathcal C_{ij}=\langle X_iX_j\rangle=\mathrm{Tr}\hat{\rho}
\{\hat{X}_i,\hat{X}_j\}/2\label{ccc}
\end{equation} with $\{\hat{X},\hat{Y}\}=\hat{X}\hat{Y}+\hat{Y}\hat{X}$
for arbitrary
observables $\hat{X}_i$ and density matrix $\hat{\rho}$. Such a relation is consistent with simultaneously measurable correlations.
More generally, it holds even in the noncontextual case, when observables from different settings commute.
Only these elements of the matrix $\mathcal C$ are measurable, for the rest (\ref{ccc}) is only definition.
Our construction includes
all possible first-order averages $\langle X_i\rangle$ by setting one
observable to identity or subtracting averages ($X_i\to X_i-\langle X_i\rangle$).
Since $\mathrm{Tr}\hat{\rho}\hat{W}^2\geq 0$ for
$\hat{W}=\sum_i\lambda_i\hat{X}_i$ with arbitrary real $\lambda_i$, we find that the correlation matrix $\mathcal C$ is positive definite.
Therefore every correlation can be simulated by a
classical Gaussian distribution $\varrho\propto \exp(-\sum_{ij}{{\mathcal C}^{-1}}_{ij}X_iX_j/2)$,
with $\mathcal C^{-1}$ being the matrix inverse of $\mathcal C$.
This is a LHV model reproducing all measurable correlations.
We recall that we do not assume dichotomy $X=\pm 1$, which is equivalent to $\langle
(X^2-1)^2\rangle=0$ and requires $\langle X^4\rangle$. 
For simplicity, from now on we shall fix $\langle X_i\rangle=0$, redefining all quantities $X_i\to X_i-\langle X_i\rangle$.

It is interesting to note that  Tsirelson's bound \cite{cir} can be seen as consequence of weak positivity.
Taking observables $A_1$, ${A}_2$, ${B}_1$, and ${B}_2$, we have
\begin{equation}
\langle(\sqrt{2}A_1-B_1-B_2)^2\rangle+\langle(\sqrt{2}A_2-B_1+B_2)^2\rangle\geq 0
\end{equation}
for the Gaussian distribution with the correlation matrix (\ref{ccc}).
It is equivalent to
\begin{eqnarray}
&&\langle A_1 B_1\rangle+\langle A_1 B_2\rangle+\langle A_2 B_1\rangle
-\langle A_2 B_2\rangle\nonumber\\
&&\leq(\langle A^2_1\rangle+\langle A^2_2\rangle+\langle B^2_1\rangle+\langle B^2_2\rangle)/\sqrt{2}.
\end{eqnarray}
For $A,B=\pm 1$, the right hand side gives Tsirelson's bound $2\sqrt{2}$ which is at the same time the maximal quantum value of the left-hand side.
On the other hand, the upper classical bound in this case is $2$ \cite{bell}, but it requires assuming dichotomy or  equivalently knowledge of higher moments.

\section{Third Moments}
\label{threem}

Having learned that second moments do not show nonclassicality at all, we turn to third moments.
If the matrix $\mathcal C$ is strictly positive definite, all third order correlations can be explained
by a positive probability as well (the proof in Appendix A).
The problematic case is a semipositive-definite $\mathcal C$, with at least one $0$ eigenvalue. 
One cannot violate noncontextuality with an arbitrary state and third-order correlations. To see this, let us take the completely random state $\hat{\rho}\propto\hat{1}$ and suppose that the correlation matrix (\ref{ccc}) 
has a zero eigenvalue for $\hat{W}=\sum_k\lambda_k\hat{X}_k$. Then $\langle W^2\rangle=0$ and $\mathrm{Tr}\hat{W}^2=0$, which
gives $\hat{W}=0$. We can simply eliminate one of observables by the substitution $\hat{X}_m=-\sum_{k\neq m}\lambda_k\hat{X}_k/\lambda_m$
using the symmetrized order of the operators when noncommuting products appear. Now the remaining correlations matrix $C_{ij}$ with $i,j\neq m$ is positive definite and the proof in Appendix A holds.
If the correlation matrix has more zero eigenvalues, we repeat the reasoning, until only nonzero eigenvalues remain.
Furthermore, third-order correlations alone cannot show noncontextuality in a state-dependent way for up to 4 observables, nor in any two-dimensional Hilbert space, nor they can violate local realism (proofs in Appendices B and C). There exists, however, an example of violation of state-dependent noncontextuality with five observables in three-dimensional space (Appendix D).

Instead, here we show a simple example violating state-dependent noncontextuality, based on the Greenberger-Horne-Zeilinger (GHZ) idea \cite{ghz}.
We consider a three qubit Hilbert space with the 8 basis states are denoted $|\epsilon_1\;\epsilon_2\;\epsilon_3\rangle$ with $\epsilon_\alpha=\pm$. 
We have three sets of Pauli matrices $\hat{\sigma}^{(\alpha)}_j$, with $\hat{\sigma}_1=|-\rangle\langle +|
+|+\rangle\langle -|$ and $\hat{\sigma}_2=i|-\rangle\langle +|-i|+\rangle\langle -|$,
acting only in the respective Hilbert space of qubit $\alpha$. Now let us take
the six observables,
$\hat{A}_\alpha=\hat{\sigma}^{(\alpha)}_1$ , $\hat{B}_\alpha=\hat{C}\hat{\sigma}^{(\alpha)}_2$ for $\alpha=1,2,3$ and
$\hat{C}=\hat{\sigma}^{(1)}_2\hat{\sigma}^{(2)}_2\hat{\sigma}^{(3)}_2$. 
All $\hat{A}$'s commute with each other, similarly all $\hat{B}$'s commute, and $\hat{A}_\alpha$ commutes with $\hat{B}_\alpha$. 
We take  $\hat{\rho}=|\mathrm{GHZ}\rangle\langle \mathrm{GHZ}|$ for the GHZ state
\begin{equation}
\sqrt{2}|\mathrm{GHZ}\rangle=|+  +  \:+\rangle+|-  - \: -\rangle.
\end{equation}
Assuming noncontextuality, we have
\begin{equation}
\langle(A_\alpha+B_\alpha)^2\rangle=\mathrm{Tr}\hat{\rho}(\hat{A}_\alpha+\hat{B}_\alpha)^2=0,
\end{equation}
which implies $A_\alpha=-B_\alpha$, so classically $\langle A_1 A_2 A_3\rangle=-\langle B_1 B_2 B_3\rangle$.
However,
\begin{eqnarray}
&&\langle A_1 A_2 A_3\rangle=\mathrm{Tr}\hat{\rho}\hat{A}_1\hat{A}_2\hat{A}_3=1,\nonumber\\
&&\langle B_1 B_2 B_3\rangle=\mathrm{Tr}\hat{\rho}\hat{B}_1\hat{B}_2\hat{B}_3=1,
\end{eqnarray}
in contradiction with the earlier statement and excluding noncontextual LHVs. 
Hence, we have seen that the third order correlations may violate noncontextuality for specific states. It should  not be surprising that the test is based on violating an equality, instead of an inequality, because third moments can have arbitrary signs.


\section{Fourth-order correlations: noncontextuality}

To find a test of noncontextuality we now consider fourth moments. Mermin and Peres \cite{merper} have shown a beautiful example of state-independent violation of noncontextuality using observables on the tensor product
of two two-dimensional Hilbert spaces $\mathcal H_A\otimes \mathcal H_B$ arranged in a square
\begin{equation}
\begin{array}{|c|c|c|c|}
\hline
\hat{M}_{ij}&j=1&j=2&j=3\\
\hline
i=1&\hat{\sigma}^A_1&\hat{\sigma}^A_{1}\hat{\sigma}^B_1&\hat{\sigma}^B_1\\
\hline
i=2&-\hat{\sigma}^A_1\hat{\sigma}^B_3&\hat{\sigma}^A_2\hat{\sigma}^B_2&-\hat{\sigma}^A_3\hat{\sigma}^B_1\\
\hline
i=3&\hat{\sigma}^B_3&\hat{\sigma}^A_3\hat{\sigma}^B_3&\hat{\sigma}^A_3\\
\hline
\end{array}
\label{mpp}
\end{equation}
where the Pauli observables $\hat{\sigma}_i$ are in each Hilbert space ($\{\hat{\sigma}_i,\hat{\sigma}_j\}=2\delta_{ij}\hat{1}$).
Observables in each row and each column commute. We denote products in each column $\hat{C}_i=\hat{M}_{1i}\hat{M}_{2i}\hat{M}_{3i}$ and
row $\hat{R}_i=\hat{M}_{i1}\hat{M}_{i2}\hat{M}_{i3}$. We get $\hat{C}_i=-\hat{1}$ and $\hat{R}_i=\hat{1}$. If $\hat{M}_{ij}$ are replaced by classical variable $M_{ij}$ then $C_1C_2C_3=R_1R_2R_3$ in contradiction with the quantum result. 

Now we assume that the $M$ are not spin-$1/2$, but arbitrary operators, which can grouped into a Mermin-Peres square fulfilling the corresponding commutation relations, $[\hat{M}_{ij},\hat{M}_{ik}]=[\hat{M}_{ij},\hat{M}_{kj}]=0$ (operators in the same column or row commute).
We will show that in this example the dichotomy test can be avoided by fourth-order correlations, without other assumptions on values $M_{ij}$. To see this, 
note that 
$
S\equiv \sum_i (C_i-R_i)=\det N,
$
where $N_{ij}= M_{i+j,i-j}$ (counting modulo 3). Now, we note that $(\det N)^2=\det(N^TN)$
and  the eigenvalues $\lambda_i$ of $N^TN$ are real and positive. Using the Cauchy inequality 
we find that $\det(N^TN)=\lambda_1\lambda_2\lambda_3\leq (\lambda_1+\lambda_2+\lambda_3)^3/27=(\mathrm{Tr}N^TN)^3/27$.
We get then
\begin{equation}
3\sqrt{3}|S|\leq\left(\sum\nolimits_{ij}M_{ij}^2\right)^{3/2}\leq 3\sum\nolimits_{ij}|M_{ij}|^3
\end{equation}
where we used the H\"older inequality in the last step. Now, we take the average of the above equation, use $|\langle S\rangle|\leq \langle |S|\rangle$ and apply the Cauchy-Bunyakovsky-Schwarz inequality $\langle |xy|\rangle\leq(\langle x^2\rangle
\langle y^2\rangle)^{1/2}$ to $x=M_{ij}$ and $y=M_{ij}^2$. We obtain finally an inequality obeyed by all noncontextual theories
\begin{equation}
|\langle S\rangle|\leq\sum\nolimits_{ij}\left[\langle M_{ij}^2\rangle\langle M_{ij}^4\rangle/3\right]^{1/2}. \label{ccc1}
\end{equation}
The inequality involves maximally fourth-order correlations and every correlation is measurable (corresponds to commuting observables).
One can check that if $M_{ij}$ correspond to (\ref{mpp}) then the left-hand side of (\ref{ccc1}) is $6$ while
the right-hand side of (\ref{ccc1}) is $3\sqrt{3}$, giving a contradiction. Hence, a violation of (\ref{ccc1}) is possible, but it remains to be shown that systems with naturally continuous variables violate are contextual by violating Eq.~(\ref{ccc1}) or other fourth-moment-based inequalities.

\section{Fourth-order correlations: nonlocality}

A simple fourth-moment-based inequality testing local realism has been considered by CFRD \cite{cfrd}
\be
\langle A_1B_1-A_2B_2\rangle^2+\langle A_1B_2+A_2B_1\rangle^2\leq \langle(A_1^2+A_2^2)(B_1^2+B_2^2)\rangle.\label{class}	
\ee
Note that all averages involve only simultaneously measurable quantities. This constitutes an inequality,
which holds classically, involves only 4th-order averages and is scalable with respect to $A$ and $B$.
Unfortunately,  (\ref{class}) and its generalizations \cite{schvo} are not violated at all in quantum mechanics as shown in \cite{salles}. We present an alternative
proof in Appendix E.

Unfortunately a violable two-party fourth-order inequality is much more complicated \cite{bb11}. A different, but quadripartite inequality can be obtained by a slight modification of CFRD inequalities \cite{cfrd}. It reads 
\begin{equation}
|\langle ABCD\rangle|^2\leq \langle |AB|^2\rangle\langle|CD|^2\rangle\label{ineq2a}
\end{equation}
where $A=A_1+iA_2$ etc., so that both sides, when expanded, contain only simultaneously measurable correlations
(because $|\langle ABCD\rangle|^2=\langle\mathrm{Re}ABCD\rangle^2+\langle\mathrm{Im}ABCD\rangle^2$ is free from products $\langle A_1A_2\cdots\rangle$ and $|A|^2=A_1^2+A_2^2$ on the right-hand side)
It follows from the generalized triangle inequality $|\langle Z\rangle|\leq \langle |Z|\rangle$ for $Z=ABCD$ and
the Cauchy-Bunyakovsky-Schwarz inequality $\langle XY\rangle^2\leq \langle X^2\rangle\langle Y^2\rangle$  for $X=|AB|$ and $Y=|CD|$. See more details in Appendix F.

Interestingly, the inequality (\ref{ineq2a}) can be violated by correlations of positions and momenta,
Let us take standard harmonic oscillator operators $\sqrt{2}\hat A=\hat X_A+i\hat P_A$ with $[\hat X_A,\hat P_A]=i$ ($\hbar=1$) so $A_1\to \hat X_A/\sqrt{2}$, $A_2 \to \hat P_A/\sqrt{2}$,
and $[\hat A,\hat A^\dag]=1$
and analogously for $B$, $C$, and $D$. In the Fock basis $\hat A|n\rangle_A=\sqrt{n}|n-1\rangle_A$ etc.
Now take a specific entangled state in the product space of $A$,$B$,$C$, and $D$,
$
|\psi\rangle=\sum_{n\geq 0}^{N} z_n|nnnn\rangle
$
with real $z_n$ (for simplicity) and check if (\ref{ineq2a}) holds also quantum mechanically. 
We find that $\langle\psi| \hat A\hat B\hat C\hat D|\psi\rangle=\sum_n n^2z_nz_{n-1}$ while
$
\langle\psi| (\hat A^\dagger \hat A+\hat A\hat A^\dagger)(\hat B^\dagger \hat B+\hat B\hat B^\dagger)|\psi\rangle=\sum_n z_n^2(2n+1)^2
$,  and similarly for $C$ and $D$.
Due to symmetry between the oscillators, the inequality (\ref{ineq2a}) is equivalent to
$
\langle ABCD\rangle\leq \langle |AB|^2\rangle$, and the quantum mechanical prediction reads
$
\sum_{n=0}^N n^2z_nz_{n-1}\leq \sum_{n=0}^N z_n^2(n+1/2)^2
$.
This is equivalent to the positivity of  the $(N+1)\times(N+1)$ matrix $M$ with entries
$M_{nn}=(n+1/2)^2$ for $n=0,1,\dots,N$ and $M_{n,n+1}=M_{n+1,n}=-(n+1)^2/2$ for $n=0,1,\dots,N-1$, and $0$ otherwise.
However, for $N\geq 10$ we get $\det{M}<0$ so it must have a negative eigenvalue. A numerical check for $N=10$ shows that e.g. the state with the $\{z_n\}=\{0.83, 0.42, 0.27, 0.18, 0.13, 0.09, 0.07, 0.05, 0.03, 0.02, 0.01\}$ violates (\ref{ineq2a}).
The generation of the highly entangled state violation (\ref{ineq2a}) will be difficult but possible
because techniques of generation of multipartite entangled optical states already exist \cite{pan}.

\section{Conclusions} 

We have proved that one cannot show nonclassicality by violating inequalities containing only up to third-order correlations, except state-dependent contextuality.
Fourth order correlations are sufficient to violate locality and state-independent noncontextuality but
the corresponding inequalities are quite complicated.
 A fourth order quadripartite Bell-type inequality
(\ref{ineq2a}) can be violated by 4th-order correlations of position and momentum or quadratures for special entangled states.

\section*{Acknowledgments}
We are grateful for discussions with N. Gisin and M. Reid.  A. B. acknowledges financial support by the Polish MNiSW grant IP2011  002371
W. Bednorz acknowledges partial financial
support by the Polish MNiSW Grant no. N N201 608740.
W. Belzig acknowledges financial support by the DFG via SPP 1285 and SFB 767.

\section*{Appendix A. Positive definite correlations} 
\renewcommand{\theequation}{A.\arabic{equation}}
\setcounter{equation}{0}

Let us assume that the correlation matrix $\mathcal C$ from (\ref{ccc}) is strictly positive definite, having all eigenvalues positive. We will prove that every third order correlation can explained also by a positive probability.
We also shift all first order averages to zero, $X_i\to X_i-\langle X_i\rangle$.
So far the distribution of $X$ was Gaussian and $X$ were continuous, but in this case all central third moments are zero.
To allow for nonzero third moments we have to change the probability. The simplest (but not the only) way is to
change the probability at particular values of $X$ to get a non-Gaussian distribution.
We define additional labels  $\{ijkq\}$, $i\neq j\neq k\neq i$ (in this case one for all possible permutations of $ijk$),
$\{ijq\pm \}$, $i\neq j$, $\{ijq\pm\}\neq\{jiq\pm\}$ (here order matters), and $\{iq\}$ with an auxiliary parameter $q\in\{3,-1,-2\}$. 
The modified distribution reads
\begin{eqnarray}
&&
\varrho(X)=\varrho_G(X)+
\lambda^{-3}\sum_L\prod_j\delta(X_j-W_j(L)),
\nonumber\\
&&\varrho_G(X)=\frac{1-c/\lambda^3}{(2\pi)^{n/2}(\det \mathcal C)^{1/2}}e^{-\sum_{ij}\mathcal C^{-1}_{ij}X_iX_j/2},
\end{eqnarray}
where $\varrho_G$ is the "old" Gaussian (renormalized) while the second part is the sum over delta peaks
at particular points depending on the label $L$. Here $c$ is the number of all labels $L$ and $\lambda>0$ is some very large real parameter such that $c/\lambda^3<1$.
The positions of the peaks are 
\begin{eqnarray}
&&W_{i,j,k}(\{ijkq\})=q\lambda\langle X_iX_jX_k\rangle^{1/3}/\sqrt[3]{18},\label{trr}\\
&&W_{i}(\{ijq\pm\})=\pm\sqrt{2} qQ_{ij}/\sqrt[3]{18},\nonumber\\
&&W_{j}(\{ijq\pm\})=qQ_{ij}/\sqrt[3]{18},\nonumber\\
&&Q_{ij}=\frac{\lambda}{\sqrt[3]{4}}\left[\langle X^2_iX_j\rangle-\sum_{k\neq ij}\langle X_iX_jX_k\rangle\right]^{1/3},\nonumber\\
&&W_i(\{iq\})=\frac{q\lambda}{\sqrt[3]{18}}\left[\langle X^3_i\rangle-\sum_{j\neq i}\langle X^2_jX_i\rangle/2\right]^{1/3}\nonumber\\
&&W_l(\{ijkq\})=W_l(\{ijq\pm\})=W_l(\{iq\})=0,\: l\neq ijk.\nonumber
\end{eqnarray}
The cubic root is defined real for real negative arguments. 
Here $\langle X_iX_jX_k\rangle$ are the desired third moments (the argument holds even for noncommuting observables).
 Note that the special choice of $q$ results in unchanged
averages $\langle X_i\rangle$ as $3-1-2=0$ but nonzero third order averages as $3^3-1^3-2^3=18$. 
The calculation of the third moments gives exactly the desired values.
Unfortunately, it will modify the correlation matrix $\mathcal C$. However, the correction is $\sim 1/\lambda$.
The modified correlation matrix is then arbitrarily close
to ${\mathcal C}$ at $\lambda\to\infty$, so it must be positive definite and we can find the new Gaussian part in the form 
$\varrho_G\propto \exp(-\sum_{ij}{{\mathcal C}^{\prime-1}}_{ij}X_iX_j/2)$,
where the matrix $\mathcal C'$ gives the correct total second-order correlations.

The assignment (\ref{trr}) is certainly not unique, one could easily find a lot of different ones also
reproducing correctly third order correlations.
However, the bottom line is that the proof works only if $\mathcal C$ has positive signature. If some eigenvalues of $\mathcal C$ are $0$ (which occurs when a particular $X_i$ is in fact linearly dependent on the others) then ${\mathcal C}'$ 
may have a negative eigenvalue for arbitrary $\lambda$
and we cannot find any Gaussian distribution, as shown in the example in Section \ref{threem}.

\section*{Appendix B. Noncontextuality in simple cases} 
\renewcommand{\theequation}{B.\arabic{equation}}
\setcounter{equation}{0}

Let us examine state-dependent noncontextuality with up to 4 observables, $\hat{A}_i$, $i=1,2,3,4$ with
the outcomes $A_i$ or $A,B,C,D$.
We look for a positive probability $\varrho(\{A_i\})$ that reproduces correctly all first, second and third moments
calculated by quantum rules.
We have the freedom to set values of correlations of
noncommuting products of observables because they are not measurable simultaneously. The construction of the probability depends on the commutation properties of the set $\{\hat A_i\}$ and is shown for various cases in Table \ref{tab2}.
We denote $\rho(\{A_i\})=\mathrm{Tr}\hat{\rho}\prod_i\delta(A_i-\hat{A}_i)$ for every subset of commuting $\hat{A}_i$.

\begin{table}
\begin{tabular}{|c|c|}
\hline
Observables & $\varrho(A,B,C,D)$\\
\hline
\hline
$A$ $B$ $C$ $D$ &$\rho(A)\rho(B)\rho(C)\rho(D)$\\
\hline
$A-B$ $C$ $D$&$\rho(A,B)\rho(C)\rho(D)$\\
\hline
\begin{minipage}{2cm}
\vspace{2pt}
$A$ $\:D$\\
$/\:\:\:\backslash\quad$\\
$B-C\quad$
\vspace{2pt}
\end{minipage}&$\rho(A,B,C)\rho(D)$\\
\hline
\begin{minipage}{2cm}
\vspace{2pt}
$A$ $D$\\
$B-C$
\vspace{2pt}
\end{minipage}&$\rho(A)\rho(B,C)\rho(D)$\\
\hline
$B-A-C$ $D$&$\rho(A,B)\rho(A,C)\rho(D)/\rho(A)$\\
\hline
\begin{minipage}{2cm}
\vspace{2pt}
$B-A-D$\\
$|$\\
$C$
\vspace{2pt}
\end{minipage}&$\rho(A,B)\rho(A,C)\rho(A,D)/\rho^2(A)$\\
\hline
\begin{minipage}{2cm}
\vspace{2pt}
$A-D$\\
$/\:\:\:\backslash\quad\quad$\\
$B-C\quad\quad$
\vspace{2pt}
\end{minipage}&$\rho(A,B,C)\rho(A,D)/\rho(A)$\\
\hline
\begin{minipage}{2cm}
\vspace{2pt}
$A-D$\\
$|\times|$\\
$B-C$
\vspace{2pt}
\end{minipage}&$\rho(A,B,C,D)$\\
\hline
$A-B-C-D$&$\rho(A,B)\rho(B,C)\rho(C,D)/\rho(B)\rho(C)$\\
\hline
$A-B$ $C-D$&$\rho(A,B)\rho(C,D)$\\
\hline
\begin{minipage}{2cm}
\vspace{2pt}
$A$\\
$/\:\:\:\backslash$\\
$B-C$\\
$\backslash\:\:\:/ $\\
$D$
\vspace{2pt}
\end{minipage}&$\rho(A,B,C)\rho(B,C,D)/\rho(B,C)$\\
\hline
\begin{minipage}{2cm}
\vspace{2pt}
$A_1-B_1$\\
$|\quad\quad|$\\
$B_2-A_2$
\vspace{2pt}
\end{minipage}& Appendix C\\
\hline
\end{tabular}
\caption{Construction of positive probabilities $\varrho$ for all cases of up to 4 observables. Here the link $-$ means the observables commute (not linked do not commute). Exceptions: $\varrho=0$ when the denominator is zero.}
\label{tab2}
\end{table}

The only difficult case is with noncommuting pairs $(\hat{A}_1,\hat{A}_2)$
and $(\hat{B}_1,\hat{B}_2)$  but this is equivalent to the test of local realism. We will show in the general proof that this case can be always (if we do not use fourth moments) explained by a LHV model in Appendix C. 
Thus, we have shown that it is possible to define positive probability distributions $\varrho$ that reproduces all quantum first, second, and third moments of measurable (commuting) combinations of up to 4 observables.

In two-dimensional Hilbert space the situation is somewhat simpler and we can find a classical construction for an arbitrary number of observables (not limited to 4). Observables have the structure
$\hat{A}=a_0\hat{1}+\vec{a}\cdot\hat{\vec{\sigma}}$, where $\hat{\vec{\sigma}}=(\hat{\sigma}_1,\hat{\sigma}_2,\hat{\sigma}_3)$
with standard Pauli matrices $\hat{\sigma}_j$, satisfying $\{\hat{\sigma}_j,\hat{\sigma}_m\}=2\delta_{jm}\hat{1}$.
Observables $\hat{A}$ and $\hat{B}$ commute if and only if $\vec{a}\parallel\vec{b}$. We can group all observables (their number is arbitrary) parallel to
the same direction, so that $\vec{a}_\alpha\parallel\vec{a}$, $\vec{b}_\beta\parallel\vec{b}$, $\vec{c}_\gamma\parallel\vec{c}$, $\dots$,
where $\vec{a}\nparallel\vec{b},\vec{c},\dots$, $\vec{b}\nparallel\vec{c},\dots$, etc.. Then we  construct a LHV model
defined by $\varrho(\{A_\alpha\},\{B_\beta\},\{C_\gamma\},\dots)=
\rho(\{A_\alpha\})\rho(\{B_\beta\})\rho(\{C_\gamma\})\cdots$, where $\rho(\{A_\alpha\})=\mathrm{Tr}\hat{\rho}\prod_j\delta(A_\alpha-\hat{A}_\alpha)$ and similar for the other sets. This means that
all (noncontextual) third moments for a two-level system are reproduced by a classical probability.

On the other hand we will see in Appendix D an example of the violation of state-dependent noncontextuality involving a three-dimensional Hilbert space and 5 observables.

\section*{Appendix C. Third moments -- contextual LHV models} 
\renewcommand{\theequation}{C.\arabic{equation}}
\setcounter{equation}{0}
\renewcommand{\thesubsection}{\arabic{subsection}}
We will present a general proof that third order correlations can be explained by a LHV model, if 
contextuality is allowed and no assumption on higher order moments or dichotomy is made. As in Section II, we denote
$\mathcal C_{X\alpha j,Y\beta k}=\langle X_{\alpha j}Y_{\beta k}\rangle$ for $X,Y=A,B,C,\dots$ and $\alpha,\beta,j,k=1,2,\dots$.
For a valid LHV theory, $\mathcal C$ must be positive (semi)definite.

\subsection{Assumptions}
The proof is based on two facts:
\begin{itemize}
\item  $\mathcal C_{X\alpha j,X\beta k}=\langle X_{\alpha j}X_{\beta k}\rangle$ is not measurable for $\alpha\neq\beta$ (even if accidentally
 $\hat{X}_{\alpha j}\hat{X}_{\beta k}=\hat{X}_{\beta k}\hat{X}_{\alpha j}$) because $\alpha$ and $\beta$ correspond to two different settings of the same observer which cannot be realized simultaneously. So it is a free parameter in a LHV model. 
\item  We can always redefine every observable
within one observer's setting by a real linear transformation
$\hat{X}_{\alpha m}\to \sum_k\lambda_{\alpha k}\hat{X}_{\alpha k}$ as long as the linear independence is preserved, because all such observables commute with each other.
\end{itemize}
The proof involves a kind of Gauss elimination on a set of linear equations \cite{algebra}.

\subsection{Problem of zero eigenvalues}
The first choice for $\mathcal C$ will be (\ref{ccc}), which is positive semidefinite. We shall see that this choice must be sometimes modified, 
without affecting the \emph{measurable} correlations. 
Suppose that the correlation matrix $\mathcal C$ has $\mathcal N$ zero eigenvalues with linearly independent zero eigenvectors 
\begin{equation}
W_m=\sum_{\alpha,k}^{X=A,B,...}\lambda^m_{X\alpha k}X_{\alpha k}\,,\, m=1..\mathcal N
\end{equation}
with the property $\langle W_m^2\rangle=0$. This implies $\mathrm{Tr}\hat{\rho}\hat{W}_m^2=0$, which gives 
\begin{equation}
W_m=\hat{W}_m\hat{\rho}=0,\:m=1..\mathcal N.\label{eqq}
\end{equation}
The above set of linear equations can be modified as in usual algebra, we can multiply equations by nonzero numbers and add up, as long as
the linear independence holds. Vectors $W_m$ span the kernel of the correlation matrix.
We shall prove that for a given observer $X$ the above set of equations can be written in the form
\begin{equation}
X_{\alpha k}+\sum_{\beta j}^{Y\neq X}\lambda^{X\alpha k}_{Y\beta j}Y_{\beta j}=0\label{corr}
\end{equation}
where we sum over all observers different from $X$ and all their settings and observables
plus equations not containing $X$. If this were not possible then we shall prove that we can reduce the kernel by at least one vector by modifying nonmeasurable correlations
in the correlation matrix, keeping its positivity. By such successive reduction we will end up with (\ref{corr}).
For the Bell case ($A$ and $B$, $\alpha=1,2$) (\ref{corr}) reduces either to trivial single vectors $A_\alpha-\lambda B_\alpha$ or a set
\begin{equation}
\left\{\begin{array}{ccc}
A_1&=&\lambda_{11}B_1+\lambda_{12}B_2\\
A_2&=&\lambda_{21}B_1+\lambda_{22}B_2\end{array}\right.
\end{equation}
with invertible matrix $\lambda$.
 The original correlation matrix (\ref{ccc}) may lead us into troubles for some correlations (violation of noncontextuality), which are
anyhow unobservable so we do not need to bother in contextual LHV models.
Therefore, sometimes we have to modify it slightly to relax dangerous constraints. The resulting LHV correlation matrix can be different
from (\ref{ccc}) but only for nonmeasurable correlations. We make use of the fact that quantum mechanics does not permit to measure everything
in one run of the experiment, leaving more freedom for contextual LHV models.

\subsection{Reduction of zero eigenvectors}
We shall prove that all zero eigenvectors can be eliminated except those in the form of (\ref{corr}).
Without loss of generality let us take $X=A$.
We write (\ref{eqq}) in the form
\begin{equation}
\sum_{\alpha k}\lambda_{\alpha k}^m A_{\alpha k}+\nA=0,\label{sss}
\end{equation}
where $\nA$ replaces all linear combinations of quantities measured by the other observers ($B$, $C$, $D$, ...), e.g.
$\nA$ can be $2B_{11}-3B_{1 1}+B_{21}-5C_{1 3}$.
By linear eliminations and transformations within setting $1$, there exists a form of (\ref{sss}) consisting of
\begin{equation}
A_{1 k}+\naa +\nA=0,\:k=1,2,\dots,\label{eee}
\end{equation}
with $\naa$ not containing $A_{1j}$ terms, and other equations that do not contain $A_{1j}$ at all. 
Suppose that at least one of (\ref{eee}) contains an $A_{2j}$ term, so in general (\ref{eee}) has the form
\begin{equation}
A_{1 k}+\sum_m\lambda_{km} A_{2 m}+\naa\nbb +\nA=0,\:k=1,2,\dots\label{eeb}
\end{equation}
with at least one $\lambda_{km}\neq 0$ and $\naa\nbb$ denoting all terms not containing $A_{1j}$ and $A_{2j}$.
By linear eliminations and transformations within settings $1$ and $2$ we arrive at
\begin{eqnarray}
&&A_{1 k}+ A_{2 k}+\naa\nbb +\nA=0,\:k=1,2,\dots, l\nonumber\\
&&A_{1 k}+\naa\nbb +\nA=0,\:k=l+1,l+2,\dots,
\label{eeb1}\\
&&A_{2 k}+\naa\nbb +\nA=0,\:k=l+1,l+2,\dots,\nonumber
\end{eqnarray}
and other equations that do not contain $A_{1j}$ nor $A_{2j}$ at all (if we have a single observable for each setting then we can omit the index $k$). If $l>0$ then we change 
$\langle A_{1 1}A_{2 1}\rangle\to
\langle A_{1 1}A_{2 1}\rangle+\epsilon$ with $\epsilon>0$ in the correlation matrix $\mathcal C$ (or $\langle A_1A_2\rangle$ for single observables). Then $\langle W^2\rangle=2\epsilon>0$,
where $W$ is the left hand side of the first line in (\ref{eeb1}) for $k=1$. Correlations involving other kernel vectors remain unaffected as none of them
contains $A_{11}$ nor $A_{21}$. For sufficiently small, but positive $\epsilon$ the new correlation matrix $\mathcal C$ will be strictly positive
for in the space spun by the old non-kernel vectors plus $W$. In this way we reduce by $1$ the dimension of the kernel. 
By repeating this reasoning we kick out of the kernel all vectors on the left hand side of the first line of (\ref{eeb1}).
Once we are left with only two last lines of (\ref{eeb1}) we proceed by induction.

Let us assume that, at some stage with a fixed $\alpha$, the kernel equations have the form
\begin{equation}
A_{\xi k}+\sum_{m}\lambda^\xi_{km} A_{\alpha m}+\naa\cdots\nal +\nA=0\label{eeo1}
\end{equation}
for all $\xi<\alpha$ plus other equations not containing $A_\xi$ and $A_{\alpha m}$.
Note that the set of possible $k$ can be different for different $\xi$.
If all $\lambda=0$ then we can proceed to the next induction step, taking next setting.
Otherwise, let us denote by $\Xi$ the set of all $\xi$ with $\lambda^\xi_{k1}\neq 0$ for some $k$ (we fix the other index to $1$ without loss of generality).
By linear eliminations we find only one such $k$ for each $\xi\in\Xi$ so that $\lambda^\xi_{k1}=\delta_{k1}$.
Now, we make a shift of the nonmeasurable correlations
$\langle A_{\xi 1}A_{\alpha 1}\rangle\to\langle A_{\xi 1}A_{\alpha 1}\rangle+\epsilon$ and 
$\langle A_{\xi 1}A_{\eta 1}\rangle\to\langle A_{\xi 1}A_{\eta 1}\rangle-2\epsilon$ for $\xi,\eta\in\Xi$ with $\epsilon>0$.
Denoting by $W_\xi$, $\xi\in\Xi$, subsequent left hand sides of (\ref{eeo1}) for $k=1$, we have $\langle W_\xi W_\eta\rangle
=2\epsilon\delta_{\xi\eta}$. Correlations with other kernel vectors remain zero as they do neither contain $A_{\xi 1}$ 
nor $A_{\alpha 1}$.
For sufficiently small $\epsilon$ (every new $\epsilon$ is much smaller than all previous ones), 
the correlation matrix $\mathcal C$ on old non-kernel vectors plus $W_\xi$ is
strictly positive, similarly as in (\ref{eeb1}). Hence, we kick $W_\xi$ out of the kernel.
Repeating this step for subsequent $m$ we get rid of all unwanted kernel vectors and can proceed with the induction step.
Then we repeat it for each observer to finally arrive at the desired form (\ref{corr}).

\subsection{Construction of third moments}

Now, we \emph{define} all third order correlations, including noncommuting observables.
We divide all observables into two families: $V_j$ -- appearing in (\ref{corr}) and $Y_m$ -- the rest.
Now,
\begin{eqnarray}
&&\langle Y_mY_nY_p\rangle=\sum_{\sigma(mnp)}\mathrm{Tr}\hat{\rho}\hat{Y}_m\hat{Y}_n\hat{Y}_p/6,\nonumber\\
&&\langle V_jY_mY_n\rangle=\mathrm{Tr}\hat{\rho}\{\hat{V}_j,\{\hat{Y}_m,\hat{Y}_n\}\}/4,\nonumber\\
&&\langle V_kV_lY_n\rangle=\mathrm{Tr}\hat{\rho}(\hat{V}_j\hat{Y}_n\hat{V}_k+\hat{V}_k\hat{Y}_n\hat{V}_j)/2,\label{ddd}\\
&&\langle V_jV_kV_l\rangle=\sum_{\sigma(jmn)}\mathrm{Tr}\hat{\rho}\hat{V}_j\hat{V}_k\hat{V}_l/6,\nonumber
\end{eqnarray}
where $\sigma$ denotes all 6 permutations. The above definition is consistent with projective measurement for all measurable correlations.

We have to check if $\langle WZZ'\rangle=0$ for $W$ given by an arbitrary linear combination of left hand sides of (\ref{corr})
and $Z,Z'=V_j,Y_m$. If $Z,Z'=Y_m,Y_n$ it is clear because 
\begin{equation}
\hat{W}\hat{\rho}=0.\label{xrr}
\end{equation}
If $Z=Y_m$, $Z'=V_j$, then
\begin{equation}
2\langle WY_mV_j\rangle=\mathrm{Tr}\hat{\rho}(\hat{W}\hat{Y}_m\hat{V}_j+\hat{V}_j\hat{Y}_m\hat{W})=0
\end{equation}
again because of (\ref{xrr}). Finally, we need to consider $Z=V_j$, $Z'=V_k$. Because of (\ref{xrr}), we get
\begin{equation}
6\langle WV_jV_k\rangle=\mathrm{Tr}\hat{\rho}(\hat{V}_j\hat{W}\hat{V}_k+\hat{V}_k\hat{W}\hat{V}_j)\,.\label{xxr}
\end{equation}
Without loss of generality we only need to consider two cases. The first one is $V_j=A_j$, $V_k=B_k$.
If $W$ does not contain $A$ or $B$ then we can move it to the left or right and (\ref{xxr}) vanishes due (\ref{xrr}).
Now suppose $W$ contains $A_m$. By virtue of (\ref{corr}) we can write
\begin{equation}
W=A_m+\sum_n \lambda_n B_n+\nA\nB,
\end{equation}
where $\nA\nB$ denotes all terms not containing $A$ and $B$.
Moving $A_m$ and $\sum_n \lambda_n B_n+\nA\nB$ in opposite direction in (\ref{xxr}), it can be transformed into
\begin{eqnarray}
&&\mathrm{Tr}\hat{\rho}(\hat{A}_j\hat{W}\hat{B}_k+\hat{B}_k\hat{W}\hat{A}_j)=
\mathrm{Tr}\hat{\rho}(\hat{A}_j\hat{B}_k\hat{A}_m+\hat{A}_m\hat{B}_k\hat{A}_j)
\nonumber\\
&&+\mathrm{Tr}\hat{\rho}\left[\left(\sum\nolimits_n\lambda_n\hat{B}_n+\hnAB\right)\hat{A}_j\hat{B}_k\right.\nonumber\\
&&\left.+\hat{B}_k\hat{A}_j\left(\sum\nolimits_n\lambda_n\hat{B}_n+\hnAB\right)\right]\nonumber\\
&&
=\mathrm{Tr}\hat{\rho}(\hat{A}_j\hat{B}_k\hat{W}+\hat{W}\hat{B}_k\hat{A}_j),\nonumber
\end{eqnarray}
where we used the commutation rule $\hat{A}_j\hat{B}_k=\hat{B}_k\hat{A}_j$. The last expression vanishes due to (\ref{xrr}).
If $W$ contains $B_m$, we proceed analogously.

The last case is $V_j=A_j$, $V_k=A_k$. If $W$ does not contain any $A$ terms then we can move $W$ to the left or right and
(\ref{xxr}) vanishes due to (\ref{xrr}). The remaining cases, due to (\ref{corr}), have the form
$
W=A_m+\nA$
and (\ref{xxr}) reads
\begin{eqnarray}
&&\mathrm{Tr}\hat{\rho}(\hat{A}_j\hat{W}\hat{A}_k+\hat{A}_k\hat{W}\hat{A}_j)=
\mathrm{Tr}\hat{\rho}(\hat{A}_j\hat{A}_m\hat{A}_k+\hat{A}_k\hat{A}_m\hat{A}_j)\nonumber\\
&&+\mathrm{Tr}\hat{\rho}(\hnA\hat{A}_j\hat{A}_k+\hat{A}_k\hat{A}_j\hnA).\label{jkm}
\end{eqnarray}
Now we remember that (\ref{corr}) must contain also
$
W'=A_k-\nA'$
so $\hat{A}_k\hat{\rho}=\hnA'\hat{\rho}$ which gives
\begin{eqnarray}
&&\mathrm{Tr}\hat{\rho}(\hat{A}_j\hat{A}_m\hat{A}_k+\hat{A}_k\hat{A}_m\hat{A}_j)=\mathrm{Tr}\hat{\rho}(\hat{A}_j\hat{A}_m\hnA'+\hnA'\hat{A}_m\hat{A}_j)\nonumber\\
&&=\mathrm{Tr}\hat{\rho}(\hnA'\hat{A}_j\hat{A}_m+\hat{A}_m\hat{A}_j\hnA')\nonumber\\
&&
=\mathrm{Tr}\hat{\rho}(\hat{A}_k\hat{A}_j\hat{A}_m+\hat{A}_m\hat{A}_j\hat{A}_k),
\end{eqnarray}
so (\ref{jkm}) reads
$
\mathrm{Tr}\hat{\rho}(\hat{W}\hat{A}_j\hat{A}_k+\hat{A}_k\hat{A}_j\hat{W})$
which vanishes due to (\ref{xrr}). We see that correlations containing arbitrary combinations of left hand sides of (\ref{corr}) vanish.
Now, we can simply eliminate one observable from each kernel equation (\ref{corr}), $\sum_k\lambda_k Z_k=0$,
by substitution $Z_m=-\sum_{k\neq m} \lambda_k Z_k/\lambda_m$ so that only $Z_k$, $k=1..l$ remain as independent observables.
Hence, the correlation matrix $\mathcal C$ is strictly positive (kernel is null) and we construct
the final LHV model reproducing all measurable quantum first, second and third order correlations as in Section A.
The third order correlations involving substituted observables are reproduced by virtue of the just-shown property of (\ref{ddd}).
This completes the proof.

\section*{Appendix D.Violation of state-dependent noncontextuality with third moments} 
\renewcommand{\theequation}{D.\arabic{equation}}
\setcounter{equation}{0}

There exists a third moment-based state-dependent example violating noncontextuality with 5 observables in a three-dimensional Hilbert space, which we will construct now.
Let us take observables $\hat{A}_\alpha$, for $\alpha=1,2,3,4,5$. Below all summations are over the set $\{1,2,3,4,5\}$ and indices are
counted modulo 5, $\alpha+5\mu\equiv\alpha$ with integer $\mu$. We assume that $\hat{A}_\alpha\hat{A}_{\alpha+2}=\hat{A}_{\alpha+2}\hat{A}_\alpha$ but
$\hat{A}_\alpha\hat{A}_{\alpha+1}\neq\hat{A}_{\alpha+1}\hat{A}_\alpha$, so there are 5 commuting pairs and 5 noncommuting pairs. 
Suppose that an experimentalist measures
\begin{eqnarray}
&&\mathcal S=\left\langle\left(\sum_\alpha A_\alpha\cos\frac{4\pi \alpha}{5}\right)^2\right\rangle+
\left\langle\left(\sum_\alpha A_\alpha\sin\frac{4\pi \alpha}{5}\right)^2\right\rangle\nonumber\\
&&+\left\langle\left(\sum_\alpha A_\alpha\right)^2\right\rangle\cos\frac{\pi}{5}=\sum_\alpha \langle A_\alpha^2\rangle(1+\cos(\pi/5))\nonumber\\
&&+\sum_\alpha 2\langle A_\alpha A_{\alpha+2}\rangle(\cos(\pi/5)+\cos(2\pi/5)).\label{five}
\end{eqnarray}
Let us denote
 Fourier operators $\hat{A}(q)=\sum_\alpha \hat{A}_\alpha e^{2\pi i \alpha q/5}$. Since 
$\hat{A}_\alpha=\hat{A}^\dag_\alpha$,
we have $\hat{A}(0)=\hat{A}^\dag(0)$, $\hat{A}(-1)=\hat{A}(4)=\hat{A}^\dag(1)$, $\hat{A}(-2)=\hat{A}(3)=\hat{A}^\dag(2)$.
Similarly, for outcomes $A(0)=A^\ast(0)$, $A(-1)=A(4)=A^\ast(1)$ and $A(-2)=A(3)=A^\ast(2)$ (there are either 5 real random variables or
1 real and 2 complex). We can write (\ref{five}) in the equivalent form
\begin{equation}
\mathcal S=\langle|A(2)|^2\rangle+\langle (A(0))^2\rangle\cos(\pi/5).
\end{equation}
If $\mathcal S=0$ then $A(0)=A(2)=0$. Let us further take
\begin{equation}
\mathcal Q=25\sum_\alpha\langle A_\alpha^3\rangle=\sum_{q,p,r}^{q+p+r\equiv 0}\langle A(q)A(p)A(r)\rangle.\label{three}
\end{equation}
Each term of the expansion of the right hand side must contain $A(\pm 2)$ or $A(0)$ because $\pm 1\pm 1\pm 1\not\equiv 0$ so $\mathcal S=0$ implies $\mathcal Q=0$.

Denoting the commutator by
$[\hat{X},\hat{Y}]=\hat{X}\hat{Y}-\hat{Y}\hat{X}$, we have
\begin{eqnarray}
0&=&5\sum_\alpha [\hat{A}_\alpha,\hat{A}_{\alpha+2}]e^{2\pi i\alpha q/5}\nonumber\\
&=&\sum_p[\hat{A}(q-p),\hat{A}(p)]e^{-4\pi ip/5}\nonumber\\
&=&\label{fff}\sum_p[\hat{A}(p+q),\hat{A}^\dag(p)]e^{4\pi ip/5}\,.
\end{eqnarray}
By inverse Fourier transform, satisfying the above relations for $q=1..5$ is equivalent to $[\hat{A}_\alpha,\hat{A}_{\alpha+2}]=0$.
In fact, there are only three independent equations in (\ref{fff}) for $q=0,1,2$ because $q=3,4$ can be obtained from Hermitian conjugation
of $q=2,1$ with some factor. 
We obtain
\begin{eqnarray}
 [\hat{A}(1),\hat{A}^\dag(1)]\sin\frac{\pi}{5}-[\hat{A}(2),\hat{A}^\dag(2)]\sin\frac{2\pi}{5}
&=&0,\nonumber\\
\:[\hat{A}(1),\hat{A}(0)]\sin\frac{2\pi}{5}-[\hat{A}(2),\hat{A}^\dag(1)]\sin\frac{\pi}{5} 
&=&0,\label{comm}\\
\:[\hat{A}(2),\hat{A}(0)]\sin\frac{\pi}{5}-[\hat{A}^\dag(2),\hat{A}^\dag(1)]\sin\frac{2\pi}{5}
&=&0\,.\nonumber
\end{eqnarray}
In the basis $|0\rangle$, $|1\rangle$, $|2\rangle$, we take
\begin{eqnarray}
&&\hat{A}(0)=a\left(\begin{array}{rrr}0&0&0\\0&1&0\\0&0&1\end{array}\right),\:\hat{A}(2)=b\left(\begin{array}{rrr}0&0&0\\0&1&i\\0&i&-1\end{array}\right),\nonumber\\
&&
\hat{A}(1)=c\left(\begin{array}{rrr}0&1&i\\1&0&0\\i&0&0\end{array}\right),
\end{eqnarray}
with real $a$ and complex $b,c$.
We have $[\hat{A}(0),\hat{A}(2)]=\hat{A}(1)\hat{A}(2)=\hat{A}(2)\hat{A}(1)=0$, $[\hat{A}(1),\hat{A}^\dag(1)]=2|c|^2\hat{B}$,
$[\hat{A}(2),\hat{A}^\dag(2)]=4|b|^2\hat{B}$, $[\hat{A}(1),\hat{A}(0)]=ac\hat{C}$ and $[\hat{A}(2),\hat{A}^\dag(1)]=-2bc^\ast\hat{C}$, where
\begin{equation}
\hat{B}=\left(\begin{array}{rrr}0&0&0\\0&0&-i\\0&i&0\end{array}\right),\:
\hat{C}=\left(\begin{array}{rrr}0&1&i\\-1&0&0\\-i&0&0\end{array}\right)\,.
\end{equation}
To satisfy (\ref{comm}), we need $|c|^2=4|b|^2\cos(\pi/5)$ and $bc^\ast=-ac\cos(\pi/5)$, satisfied by $b=1$, $c=2\sqrt{\cos(\pi/5)}$, $a=-1/\cos(\pi/5)$.

Assuming noncontextuality, the quantum mechanical expectation for (\ref{five}) reads,
\begin{eqnarray}
&&\mathcal S=\sum_\alpha \mathrm{Tr}\hat{\rho} \hat{A}_\alpha^2(1+\cos(\pi/5))\nonumber\\
&&+\sum_\alpha 2\mathrm{Tr}\hat{\rho}\hat{A}_\alpha\hat{A}_{\alpha+2}(\cos(\pi/5)+\cos(2\pi/5))\\
&&=\mathrm{Tr}\hat{\rho}(\hat{A}^\dag(2)\hat{A}(2)+\hat{A}(2)\hat{A}^\dag(2)+2\hat{A}^2(0)\cos(\pi/5))/2\nonumber
\end{eqnarray}
and for (\ref{three}),
\begin{equation}
\mathcal Q=25\sum_\alpha\mathrm{Tr}\hat{\rho}\hat{A}_\alpha^3=\sum_{q,p,r}^{q+p+r\equiv 0}\mathrm{Tr}\hat{\rho}\hat{A}(q)\hat{A}(p)\hat{A}(r)\,.
\end{equation}
For $\hat{\rho}=|0\rangle\langle 0|$, we have $\hat{A}(0,\pm 2)\hat{\rho}=\hat{\rho}\hat{A}(0,\pm 2)=0$, so
$\mathcal S=0$.
By explicit calculation we find,
\begin{eqnarray}
&&\mathcal Q=\langle 0|\hat{A}(1)\hat{A}(0)\hat{A}^\dag(1)|0\rangle+\langle 0|\hat{A}^\dag(1)\hat{A}(0)\hat{A}(1)|0\rangle\nonumber\\
&&+\langle 0|\hat{A}(1)\hat{A}^\dag(2)\hat{A}(1)|0\rangle+\langle 0|\hat{A}^\dag(1)\hat{A}(2)\hat{A}^\dag(1)|0\rangle\nonumber\\
&&=4a|c|^2+8\mathrm{Re}(b^\ast c^2)=8(\sqrt{5}-1)\simeq 9.9,
\end{eqnarray}
in clear contradiction to the classical prediction $\mathcal Q=0$.

\section*{Appendix E. No-go theorem on two-party CFRD inequalities} 
\renewcommand{\theequation}{E.\arabic{equation}}
\setcounter{equation}{0}

The simple fourth order CFRD-type inequalities can be constructed for two observers $A$ and $B$, with up
to 8 settings (and a single real outcome for each setting) \cite{cfrd,schvo}, 
$A^{r/i}_\alpha,B^{r/i}_\alpha$ with $\alpha={0,1,2,3}$, and read
\ba
&&\nonumber|\langle A_0B^\dag_0+A_1B^\dag_1+A_2B^\dag_2+A_3B^\dag_3\rangle|^2\\
&&+|\langle A_0B_1-A_1B_0+A^\dag_2B^\dag_3-A^\dag_3B^\dag_2\rangle|^2+\nonumber\\
&&|\langle A_0B_2-A_2B_0+A^\dag_3B^\dag_1-A^\dag_1B^\dag_3\rangle|^2\nonumber\\
&&+|\langle A_0B_3-A_3B_0+A^\dag_1B^\dag_2-A^\dag_2B^\dag_1\rangle|^2\leq\label{class2} \\
&&\sum_{\alpha\beta}\langle(A^\dag_\alpha A_\alpha+A_\alpha^\dag A_\alpha)
(B^\dag_\beta B_\beta+B_\beta^\dag B_\beta)\rangle/4,\nonumber
\ea
where we have denoted $C=C^r+iC^i$, $C=A_\alpha,B_\alpha$.
The notation is the same in classical and quantum case except 
$\hat{}$ and ${}^\dag\to{}^\ast$.
We use the complex form only to save space but
all the inequality can be expanded into purely real terms \cite{schvo}.
The inequality reduces to (\ref{class}) if we leave only $A_1^r$, $A_2^r$, $B_1^r$, $B_2^r$, while
other observables are zero.
Classically, (\ref{class2}) follows from inequality $|\langle z\rangle|^2\leq\langle |z|^2\rangle$ applied to each term on the left hand side
and summed up. Surprisingly, the inequality is not violated at all in quantum mechanics, which has been
proved in \cite{salles}.
Below we present an alternative proof.

It suffices to prove (\ref{class2}) for pure states, $\hat{\rho}=|\psi\rangle\langle\psi|$. For mixed states $\hat{\rho}=\sum_k p_k|\psi\rangle\langle\psi|$, $p_k\geq 0$, $\sum_k p_k=1$.
We apply the triangle inequality $|\sum_k p_kz_k|\leq \sum_k p_k|z_k|$ and the Jensen inequality
$(\sum_k p_k|z_k|)^2\leq \sum_k p_k|z_k|^2$, where $z_k$ is the complex correlator in each of the four terms
on the left hand side of (\ref{class2}) taken for a pure state $|\psi_k\rangle$. If (\ref{class2}) is valid for each
$|\psi_k\rangle$ then it holds for the mixture, too.

Let us focus then on pure states.
Note that the sum of the last three terms on the left hand side of (\ref{class2}) can be written as
\ba
&&\sum_{\alpha\beta}
\left(\langle A_\alpha B_\beta\rangle\langle A^\dag_\alpha B^\dag_\beta\rangle
-\langle A_\alpha B_\beta\rangle\langle A^\dag_\beta B^\dag_\alpha\rangle\right)+
\\
&&\sum_{\alpha\beta\gamma\delta}\epsilon_{\alpha\beta\gamma\delta}\left(\langle A_\alpha B_\beta\rangle
\langle A_\gamma B_\delta\rangle+\langle A^\dag_\alpha B^\dag_\beta\rangle
\langle A^\dag_\gamma B^\dag_\delta\rangle\right)/2,\nonumber
\ea
using the completely antisymmetric tensor $\epsilon$ with
$\epsilon_{0123}=1$. Therefore the whole inequality is invariant under
SU(4) transformations of $A_\alpha$, $B_\beta$ treated as components
of four-dimensional vectors (it is straightforward
to verify the invariance of other parts of the inequality). We remind that these
external transformations do not interfere with the internal Hilbert spaces 
$\mathcal H_{A,B}$. 

Let us number the four complex correlators inside the moduli on the 
left hand side of (\ref{class2}) by 
$0,1,2,3$, respectively (e.g. $0$ is the correlator $\sum_\alpha \langle A_\alpha B^\dag_\alpha\rangle$).
We want to transform (\ref{class2}) to a form
with a single real correlator $0$ while $1,2,3$ vanish. Let us begin with
a transformation $C_\alpha\to e^{i\phi_\alpha}C_\alpha$, $C=A,B$, with 
$\sum_\alpha\phi_\alpha=0$. Note that 
$A_0B_1-A_1B_0+A^\dag_2B^\dag_3-A^\dag_3B^\dag_2$  just gets the phase factor
$e^{i(\phi_0+\phi_1)}$, so tuning $\phi_\alpha$ we can always make 
the correlators $1,2,3$ real.
Making now a real rotation in $123$ space we can leave only the real correlator $3$ while 
$1$ and $2$ vanish. Still, 
the correlator $0$ can have also an unwanted imaginary component, because $0$
is invariant under SU(4) transformations. To get rid of it, 
we have to apply a different
transformation $A_0\to A_0$, $A_1\to A_1$, $A_2\to A_2^\dag$, $A_3\to A_3^\dag$,
$B_0\to-B^\dag_1$, $B_1\to B_0^\dag$, $B_2\to -B_3$, $B_3\to B_2$, which gives
\ba
&&A_0B_1-A_1B_0+A^\dag_2B^\dag_3-A^\dag_3B^\dag_2\nonumber\\
&&\to A_0B_0^\dag+A_1B_1^\dag+A_2B^\dag_2+A_3B^\dag_3\,,\nonumber\\
&&A_0B_2-A_2B_0+A^\dag_3B^\dag_1-A^\dag_1B^\dag_3\nonumber\\
&&\to-A_0B_3+A_3B_0-A_1^\dag B_2^\dag+A_2^\dag B^\dag_1\,,\label{trans}\\
&&A_0B_3-A_3B_0+A^\dag_1B^\dag_2-A^\dag_2B^\dag_1\nonumber\\
&&\to A_0B_2-A_2B_0+A^\dag_3B_1^\dag-A^\dag_1B^\dag_3\,,\nonumber\\
&&A_0B^\dag_0+A_1B^\dag_1+A_2B^\dag_2+A_3B^\dag_3\nonumber\\
&&\to-A_0B_1+A_1B_0-A^\dag_2B^\dag_3+A^\dag_3B^\dag_2\,.\nonumber
\ea
It is clear that the inequality (\ref{class2}) remains unchanged
(we can change signs in the second and fourth part of (\ref{trans})).
Now the correlator $0$ vanishes because it is moved to $-1$ and $1$ is moved to $0$ ($2\to -3$, $3\to 2$). Applying again an SU(4) transformation,
 we can get correlator $1$ real while $2,3$ vanish and $0$ remains null because
 it is invariant under SU(4). Applying again (\ref{trans}) we get only a single real term in $0$.
In this way, the left hand side
of (\ref{class2}) reads
\be
\left(\mathrm {Re}\sum_\alpha\langle A_\alpha B^\dag_\alpha\rangle\right)^2\,.
\label{ree}
\ee
We apply the triangle inequality 
\be
\left|\sum_\alpha^{q=r,i} \langle A^q_\alpha B^q_\alpha\rangle\right|\leq
\sum_\alpha^{q=r,i}|\langle A^q_\alpha B^q_\alpha\rangle|\,.
\ee
Note that $|\langle A^q_\alpha B_\alpha^q\rangle|\leq \langle |A_\alpha^q||B^q_\alpha|\rangle$
where $|X|$ is obtained by reversing signs of all negative eigenvalues of $X$ (in the eigenbasis).
To prove (\ref{class}) we have to show that
\be
\left(\sum_\alpha^{q=r,i}\langle |A^q_\alpha| |B^q_\alpha|\rangle\right)^2\leq
\sum_{\alpha\beta}^{q,p=r,i}\langle |A^q_\alpha|^2 |B^p_\beta|^2\rangle\label{in1}
\ee
We decompose $|\psi\rangle$, and arbitrary operators $\hat{A}^x$, $\hat{B}^x$ in basis $|k_A i_B\rangle$
of the joint Hilbert space $\mathcal H_A\otimes\mathcal H_B$,
\ba
&&|\psi\rangle=\sum_{ki}\psi_{ki}|k_Ai_B\rangle,\;\hat{A}^x=\sum_{kli}A^x_{kl}|k_A i_B\rangle\langle l_A i_B|,\nonumber\\
&&\hat{B}^x=\sum_{kij}B^x_{ij}|k_A i_B\rangle\langle k_A j_B|.
\ea
The normalization reads $\sum_{ki}|\psi_{ki}|^2=1$.
Let us define $\hat{\Psi}=\sum_{ki}\psi_{ki}|k\rangle\langle i|$, $\hat{a}^x=\sum_{kl}A_{kl}|k\rangle\langle l|$, $\hat{b}^x=\sum_{ij}B_{ij}|j\rangle\langle i|$. Now the normalization reads $\tr \hat{\Psi}^\dag\hat{\Psi}=1$. One can check the identity
$\langle\psi|\hat{A}^x\hat{B}^x|\psi\rangle=
\tr\hat{\Psi}^\dag\hat{a}^x\hat{\Psi}\hat{b}^x$. We stress that  $\hat{a}^x$ and
$\hat{b}^x$ 
are no longer operators in $\mathcal H_A\otimes\mathcal H_B$,
but in $\mathcal H_A$ and $\mathcal H_B$, respectively, while 
$\hat{\Psi}$ is a linear transformation from $\mathcal H_B$ to $\mathcal H_A$,
which need not be represented by a Hermitian nor even a square matrix.
We note that such a manipulation is possible only for two observers. By taking suitable bases, we could even make $\hat{\Psi}$ 
diagonal, real and positive, analogously to a Schmidt decomposition, 
but it is not necessary. Now (\ref{in1}) reads
\be
\left(\sum_\alpha^{q=r,i}\tr\hat{\Psi}^\dag|\hat{a}^q_\alpha|\hat{\Psi} |\hat{b}^q_\alpha|\right)^2\leq
\sum_{\alpha\beta}^{q,p=r,i}\tr\hat{\Psi}^\dag|\hat{a}^q_\alpha|^2\hat{\Psi} |\hat{b}^p_\beta|^2\label{in2}
\ee
To prove (\ref{in2}) we need the Lieb concavity theorem \cite{lieba}
which states that
for a fixed but arbitrary $\hat{\Psi}$ and $s\in[0,1]$ the trace class
function
$
f(\hat{F},\hat{G})=\tr\hat{\Psi}^\dag \hat{F}^s\hat{\Psi}\hat{G}^{1-s}
$
is \emph{jointly concave}, which means that
\ba
&&\lambda f(\hat{F},\hat{G})+(1-\lambda)f(\hat{F}',\hat{G}')\leq\nonumber\\
&&f(\lambda\hat{F}+(1-\lambda)\hat{F}',\lambda\hat{G}+(1-\lambda)\hat{G}')\label{lieb1}
\ea
for $\lambda\in[0,1]$ and arbitrary Hermitian semipositive operators $\hat{F},\hat{F}'$, $\hat{G},\hat{G}'$.
By induction (\ref{lieb1}) generalizes straightforward to
\be
\sum_\alpha\lambda_\alpha f(\hat{F}_\alpha,\hat{G}_\alpha)\leq
f\left(\sum_\alpha\lambda_\alpha\hat{F}_\alpha,\sum_\beta\lambda_\beta\hat{G}_\alpha\right)\label{lieb2}
\ee
for $\lambda_\alpha\geq 0$ and $\sum_\alpha \lambda_\alpha=1$ and arbitrary semipositive operators
$\hat{F}_\alpha$, $\hat{G}_\alpha$. We apply (\ref{lieb2}) for $s=1/2$, $\lambda^q_\alpha=1/8$, 
$F^q_\alpha=|a_\alpha^q|^2$ and $G^q_\alpha=|b_\alpha^q|^2$ to get
\ba
&&\sum_\alpha^{q=r,i}\tr\hat{\Psi}^\dag|\hat{a}^q_\alpha|\hat{\Psi} |\hat{b}^q_\alpha|\leq
\nonumber\\
&&\tr\hat{\Psi}^\dag\left(\sum_\alpha^{q=r,i}|\hat{a}^q_\alpha|^2\right)^{1/2}
\hat{\Psi}\left( \sum_\beta^{p=r,i}|\hat{b}^p_\beta|^2\right)^{1/2}\,.
\ea
Finally we use the operator Cauchy-Bunyakovsky-Schwarz inequality $|\tr\hat{c}\hat{d}|^2\leq\tr \hat{c}\hat{c}^\dag 
\tr \hat{d}\hat{d}^\dag$ for  $\hat{c}=\hat{\Psi}^\dag$ and
\be 
\hat{d}=\left(\sum_\alpha^{q=r,i}|\hat{a}^q_\alpha|^2\right)^{1/2}
\hat{\Psi}\left( \sum_\beta^{p=r,i}|\hat{b}^p_\beta|^2\right)^{1/2}
\ee
which completes the proof.
It is impossible to generalize CFRD inequalities to more observables \cite{schvo}.

\section*{Appendix F. Four-parties CFRD inequalities} 
\renewcommand{\theequation}{F.\arabic{equation}}
\setcounter{equation}{0}
For a complex random variable $Z$ we have a generalized triangle (in complex plane) inequality
$|\langle Z\rangle|\leq \langle |Z|\rangle$.
Now, for complex random variables $A,B,C,D$ we have
\begin{equation}
|\langle ABC\rangle|^2\leq \langle |ABC|\rangle^2\leq \langle |AB|^2\rangle\langle|C|^2\rangle\label{ineq1}
\end{equation}
and
\begin{equation}
|\langle ABCD\rangle|^2\leq \langle |ABCD|\rangle^2\leq \langle |AB|^2\rangle\langle|CD|^2\rangle,\label{ineq2}
\end{equation}
where we use Cauchy-Bunyakovsky-Schwarz inequality in the last step. Complex variables can be constructed out of real ones,
$A=A_1+iA_2$, etc., where $A_1,A_2$ are real.
Both sides of inequalities can be expanded in real variables, in such a way that no average contains simultaneously $A_1$ and $A_2$. In particular
\ba
&&\langle|AB|^2\rangle=\langle A_1^2B_1^2\rangle+\langle A_1^2B_2^2\rangle+
\langle A_2^2B_1^2\rangle+\langle A_2^2B_2^2\rangle,\nonumber\\
&&\langle |C|^2\rangle=\langle C_1^2\rangle+\langle C_2^2\rangle
\ea
while
\ba
&&|\langle ABCD\rangle|^2=\langle ABCD\rangle^\ast\langle ABCD\rangle\nonumber\\
&&=\langle\mathrm{Re}\;ABCD\rangle^2+\langle\mathrm{Im}\;ABCD\rangle^2,
\ea
where
\ba
&&\langle\mathrm{Re}\;ABCD\rangle=\langle A_1B_1C_1D_1\rangle-
\langle A_1B_1C_2D_2\rangle\nonumber\\
&&-\langle A_2B_2C_1D_1\rangle+\langle A_2B_2C_2D_2\rangle
\nonumber\\
&&-
\langle A_1B_2C_1D_2\rangle-
\langle A_1B_2C_2D_1\rangle\nonumber\\
&&-\langle A_2B_1C_1D_2\rangle-\langle A_2B_1C_2D_1\rangle
\ea
and
\ba
&&\langle\mathrm{Im}\;ABCD\rangle=\langle A_1B_1C_1D_2\rangle+
\langle A_1B_1C_2D_1\rangle\nonumber\\
&&+\langle A_1B_2C_1D_1\rangle+\langle A_2B_1C_1D_1\rangle
\nonumber\\
&&-
\langle A_1B_2C_2D_2\rangle-
\langle A_2B_1C_2D_2\rangle\nonumber\\
&&-\langle A_2B_2C_1D_2\rangle-\langle A_2B_2C_2D_1\rangle.
\ea

As quantum counterexamples, let us take spin observables $\sigma_1=|+\rangle\langle -|+|-\rangle\langle +|$
and $\sigma_2=i|-\rangle\langle +|-i|+\rangle\langle -|$.
Now $A_1=\sigma^A_1$, $A_2=\sigma^A_2$ so that $A=A_1+iA_2=\sigma^A_+=2|+\rangle\langle -|$, etc.
Taking Greenberger-Horne-Zeilinger states 
$$
\sqrt{2}|\psi\rangle=|+++\rangle+|---\rangle,\sqrt{2}|\psi\rangle=|++++\rangle+|----\rangle,
$$
we get on the left hand side of (\ref{ineq1}) 16 while the right hand side is equal to 8 and on the left hand side of (\ref{ineq2}) 64 while the right hand side is equal to 16. So in both cases they are violated.

We can test the inequalities also by position and momentum measurement.
Let us take $\sqrt{2}A=X_A+iP_A$ with $[X_A,P_A]=i$ ($\hbar=1$) so $A_1=X_A/\sqrt{2}$, $A_2=P_A/\sqrt{2}$
and $[A,A^\dag]=1$
and analogously for $B$, $C$ and $D$. In the Fock basis $A|n\rangle_A=\sqrt{n}|n-1\rangle_A$ and so on.
Now we take a generic entangled state
$$
|\psi\rangle=\sum_{n\geq 0} z_n|nnnn\rangle
$$
with real $z_n$ (for simplicity)
and check if (\ref{ineq2}) holds. 
Note that $\langle ABCD\rangle=\sum_n n^2z_nz_{n-1}$ while
\ba
&&\langle |AB|^2\rangle=\langle |CD|^2\rangle=\\
&&\langle (AA^\dag+A^\dag A)(BB^\dag+B^\dag B)\rangle/4=\sum_n z_n^2(n+1/2)^2.\nonumber
\ea
In this case, if (\ref{ineq2}) holds then also
$
\langle ABCD\rangle\leq \langle |AB|^2\rangle$ holds, which yields
\be
\sum_n n^2z_nz_{n-1}\leq \sum_n z_n^2(n+1/2)^2.
\ee
This is equivalent to the positivity of  the $(N+1)\times(N+1)$ matrix $M$ with entries
$M_{nn}=(n+1/2)^2$ for $n=0,1,\dots,N$ and $M_{n,n+1}=M_{n+1,n}=-(n+1)^2/2$ for $n=0,1,\dots,N-1$ and $0$ otherwise.
However, for $N=10$ we get $2^{22}\det{M}=-21772303951061875$ so it must have a negative eigenvalue. The numerical check shows that the minimal eigenvalue of $M$ is $\lambda_{min}=-0.00287931$ while the normalized coefficients $z_n$ read:
($z_0$,$z_1$,$z_2$,$z_3$,$z_4$,$z_5$,$z_6$,$z_7$,$z_8$,$z_9$,$z_{10}$)=(0.828979, 0.419264, 0.26503, 0.181928, 0.129563, 0.0934879, 0.0671523, 0.0471264, 0.0314302, 0.0188364, 0.00854237) which violates (\ref{ineq2}).
Note, that for larger $N$ one can get a smaller $\lambda_{min}$, e.g. $-0.093$ for $N=3000$.

Taking an analogous state
\be
|\psi\rangle=\sum_n z_n|nnn\rangle,
\ee
unfortunately one cannot violate (\ref{ineq1}) which reads in this case
\be
\left(\sum_n z_nz_{n-1}n^{3/2}\right)^2\leq \sum_n z_n^2(n+1/2)^2\sum_n z_n^2(n+1/2)\,.
\ee
One can see it from the Minkowski inequality
\be
\left(\sum_n x_ny_n\right)\leq \sum_n x_n^2\sum_n y_n^2\,,
\ee
taking $x_n=z_{n-1}\sqrt{n-1/2}$, $y_n=\sqrt{n^3/(n-1/2)}$ for $n=1,2,\dots$.
Note also that $n^3/(n-1/2)\leq (n+1/2)^2$ because $n^3\leq (n-1/2)(n+1/2)^2=(n^2-1/4)(n+1/2)=n^2-n/4+n^2/2-1/8$, which is true due to the fact that
$n^2/2-n/4-1/8\geq 0$ for $n\geq 1$.

Interestingly, in the case of the three- and four-partite CFRD inequalities, the Lieb theorem, used in Appendix E for two parties, does not prevent the violation of a classical inequality, even in the fourth-moment version. However, the violating state in the position-momentum space is quite complicated, so an open question remains whether any simpler fourth-order inequality or simpler violating state exists.

\end{document}